\documentclass[useAMS,usenatbib,usegraphicx]{mn2e}

\title[Spectroastrometry and NLTE modelling of $\beta$ CMi's disc]{Probing the properties of Be star discs with spectroastrometry and NLTE radiative transfer modelling: $\beta$ CMi\thanks{Based on observations conducted at the European Southern Observatory (ESO), Paranal, Chile as part of the programme 082.D-0140.}}
\author[H.E. Wheelwright et al.]{H.E. Wheelwright$^{1,2}$\thanks{E-mail:
hwheelwright@mpifr-bonn.mpg.de} J.E. Bjorkman$^{3}$, R.D. Oudmaijer$^{2}$, A.C. Carciofi$^{4}$,\newauthor K.S. Bjorkman$^{3}$ and J.M. Porter$^{5}$\thanks{Deceased}\\
$^{1}$Max-Planck-Institut f\"{u}r Radioastronomie, Auf dem H\"{u}gel 69,
53121 Bonn, Germany\\
$^{2}$School of Physics \& Astronomy, University of Leeds, Woodhouse Lane, Leeds LS2 9JT, UK\\
$^{3}$Department of Physics \& Astronomy, University of Toledo, MS111 2801 W. Bancroft Street, Toledo, OH 43606, USA\\
$^{4}$Instituto de Astronomia, Geof\'{i}sica e Ci\^encias Atmosf\'ericas, Universidade de S\~ao Paulo, Rua do Mat\~{a}o 1226, Cidade Universit\'aria,\\\hspace*{2mm}05508-900 S\~ao Paulo, Brazil\\
$^{5}$Astrophysics Research Institute, Liverpool John Moores University, Twelve Quays House, Egerton Wharf, Birkenhead CH41 1 LD, UK\\}
\begin{document}

\date{Accepted yyyy Month dd. Received yyyy Month dd; in original form yyyy Month dd}

\pagerange{\pageref{firstpage}--\pageref{lastpage}} \pubyear{2002}

\maketitle

\label{firstpage}

\begin{abstract}

  While the presence of discs around classical Be stars is well
  established, their origin is still uncertain. To understand what
  processes result in the creation of these discs and how angular
  momentum is transported within them, their physical properties must
  be constrained. This requires comparing high spatial and spectral
  resolution data with detailed radiative transfer modelling. We
  present a high spectral resolution, $R$$\sim$80,000, sub
  milli-arcsecond precision, spectroastrometric study of the
  circumstellar disc around the Be star $\beta$ CMi. The data are
  confronted with three-dimensional, NLTE radiative transfer
  calculations to directly constrain the properties of the
  disc. {{Furthermore, we compare the data to disc models featuring
      two velocity laws; Keperian, the prediction of the
      viscous disc model, and angular momentum conserving
      rotation}}. It is shown that the observations of $\beta$ CMi can
  only be reproduced using Keplerian rotation. {{The agreement
      between the model and the observed SED, polarisation and
      spectroastrometric signature of $\beta$ CMi confirms that the
      discs around Be stars are well modelled as viscous decretion
      discs.}}

\end{abstract}

\begin{keywords}
Physical Data and Processes: polarisation -- stars: emission-line, Be -- stars individual: $\beta$ CMi -- techniques: high angular resolution -- methods: numerical
\end{keywords}

\section{Introduction}

Be stars are non-supergiant stars with a spectral type B that either
exhibit H{\sc{i}} Balmer emission lines in their optical spectra or
have done so in the past \citep{Collins1987}. It has been thought for
some time that this Balmer line emission originates in gaseous
disc-like structures \citep{Struve1931}. The presence of rotating,
flattened material, i.e. a disc, was initially inferred from the
typical double-peaked emission line profile that is exhibited by many
Be stars. This has since been confirmed by long-baseline optical and
near-infrared interferometry which has resolved small scale discs
around several Be stars \citep[see e.g.][]{Quirrenbach1997}. However,
despite the progress made in recent years, many questions remain with
regard to the origin of these discs \citep[][]{Porter2003}.

 \smallskip

 {{Most Be stars rotate rapidly. Indeed, it appears that Be stars
     preferentially rotate at a significant fraction,$\sim$80 per cent
     or more, of the critical velocity where the centrifugal force and
     gravity are balanced \citep[][]{Porter1996,Rivinius2006}}}. This
 rapid rotation may allow material to be lifted off the surface of the
 star and into orbit, thus forming the gaseous circumstellar disc
 \citep{Struve1931}. {{However, this requires the star to rotate
     perilously close to its critical velocity, and it is not clear
     whether Be stars rotate this rapidly or not \citep[see
     e.g.][]{Townsend2004,Cranmer2005}}}. Therefore, alternative
 mechanisms such as non-radial pulsations
 \citep{Rivinius2003,Cranmer2009} and magnetic fields
 \citep[][]{Cassinelli2002,Brown2008} have been proposed as a viable
 mechanism to lift material off the central star.

\smallskip

Regardless of how material is ejected into orbit, another mechanism is
required to distribute the material throughout the disc. The viscous
decretion disc model \citep[VDDM,][]{Lee1991} features angluar
momentum (AM) transport by turbulent viscosity to lift material into
higher orbits, thereby causing the disc to grow. A necessary result of
this model is that material is in Keplerian rotation throughout the
disc.  Other models, that do not have an AM transport mechanism, such
as non-viscous magnetically confined discs and the original wind
compressed disc model \citep{Bjorkman1993}, exhibit angular momentum
conserving (AMC) kinematics. Here we directly confront these two
fundamental predictions with astrometric data and physical modelling.

\smallskip

Evidence that the discs around Be stars are well modelled as viscous
decretion discs is mounting. An example is provided by the work of
\citet{Carciofi2009}, who modelled the cyclic variability of $\zeta$
Tau with perturbations in a viscous disc. The model of
\citet{Carciofi2009} reproduces both the VLTI/AMBER observations of
$\zeta$ Tau and the temporal variations of its H$\alpha$ and
Br$\gamma$ emission and thus provides strong support for the viscous
disc scenario. As another example, Keplerian rotation, a feature of
the VDDM, is suggested by many Be emission line profiles
\citep{Hummel2000}. As a result, the Keplerian viscous disc is
currently the favoured model of Be star discs. However, as noted by
\citet{Hummel2000}, based on the line profiles alone, it is difficult
to conclusively discount alternate velocity laws.

\smallskip

A number of discs around Be stars have now been directly probed with
optical/NIR interferometry
\citep[see][]{Vakili1998,Meilland2007b,Meillandetal2007,Carciofi2009,Pott2010,Delaa2011,Meilland2011,Kraus2012,Meilland2012}. Several
of these studies also find evidence for discs in Keplerian
rotation. However, there are a few notable exceptions
\citep[see][]{Meilland2007b,Delaa2011}. Therefore, we have yet to
arrive at a complete understanding of the properties of Be star
discs. Furthermore, many of the earlier studies that employed
spectro-interferometric data were limited in terms of spectral
resolution and/or had a sparse $u,v$ coverage. These limitations have
been overcome in a few cases \citep[see
e.g.][]{Kraus2012}. Nonetheless, few studies confront spectrally and
spatially resolved data with detailed three-dimensional, NLTE
radiative transfer calculations. Here we address this issue and
confront unique high spectral resolution, sub milli-arcsecond (mas)
precision spectroastrometric data with a detailed model of Be star
discs.

\smallskip

Spectroastrometry is a complementary technique to
spectro-interferometry that offers a valuable insight into the
kinematics of unresolved structures. The technique utilises the
spatial information present in a long-slit spectrum to deliverer
spatial information with sub-mas precision \citep[see
  e.g.][]{Bailey1998a,Whelan2008}. Spectroastrometry can offer a
higher spectral resolution than spectro-interferometry and offers an
efficient way to probe circumstellar kinematics on small
scales. Therefore, such data are well suited to comparison with model
calculations.

\smallskip

We have observed a sample of Be stars with spectroastrometry. Here we
present our results on $\beta$ CMi, a B8 type star located at a
distance of 52~pc. The disc around this object was recently resolved
with the VLTI and CHARA interferometers as reported by
\citet{Kraus2012}. Using a kinematical model, these authors show that
their observations are best described assuming Keplerian rotation. Here, we
fit our high spectral resolution data and photometric
and spectro-polarimetric data with a state-of-the-art NLTE radiative
transfer model to probe the physical properties of the disc.

\section{Observations and data reduction}
\label{observations}

$\beta$ CMi was observed on the 9th of December 2008 with the UVES
spectrograph \citep{UVES} mounted on the VLT-UT2. The red arm of the
instrument was used as attention was focused upon the H$\mathrm
\alpha$ line. Observations where conducted using the 600 g/mm grating
and the H$\alpha$ filter. The MIT-LL CCD was employed and the spatial
pixel size of 0.16$\mathrm"$ ensured the seeing, which was
$\sim$0.8$\mathrm"$, was well sampled.  Observations were conducted
with a slit width of 0.5$\mathrm"$, which resulted in a spectral
resolution $\mathrm \sim80,000$ or 4~${\mathrm{km\,s^{-1}}}$. The
spectral range was $\mathrm{\sim6545}$--$\mathrm{6580 \, \rm
  \AA}$. Data were obtained at four different slit position angles,
$\mathrm{0, 90, 180\,and\,270^{\circ}}$. This is to identify, and
eliminate, any systematic artifacts in the spectroastrometric
signatures \citep[see e.g.][]{EBrannigan2006}. Spectra of a
ThAr lamp were used to wavelength calibrate the data.

\smallskip

Data reduction was conducted using IRAF and routines
written in {\sc{idl}}. Flat field frames,
which were first corrected for the average bias level, were
combined. The averaged flat field was then normalised. Finally, raw
data were corrected using an average bias frame and the normalised,
average flat frame. The total intensity long-slit spectra were then
extracted in a standard fashion.

\smallskip

Spectroastrometry was performed by fitting Gaussian profiles to the
spatial profile of the long-slit spectra at each pixel along the
dispersion axis. This allowed the centroid position to be determined
as a function of velocity. {{The continuum position exhibited a trend in
the dispersion direction which was removed by fitting a 1-D polynomial
function to it.}} The average positional spectra for anti-parallel
position angles were combined to form the average North-South (NS) and
East-West (EW) positional spectra. All such spectra were assessed
visually to exclude features only present at a single position angle
\citep[for details on the spectroastrometric technique see
e.g.][]{Oudmaijer2008,Wheelwright2010}.

\smallskip

The spectroastrometric signature of $\beta$ CMi is shown in panel d)
of Figure~\ref{res_1}. As a result of the high SNR, the average
positional precision (the rms of the centroid in the continuum) is
less than 0.2~mas. A clear spectroastrometric signature is
observed and appears as expected for a rotating disc \citep[see
e.g.][]{Oudmaijer2011}. The amplitude of the excursions in the EW and
NS directions is of order 1~mas, and allows us to derive the Position
Angle (PA) traced by the major axis of the disc of 130$\pm$5$^{\rm
  o}$. This is consistent with the polarisation angle of $\sim45
^{\rm{\circ}}$, as this is expected to perpendicular
to the disc. The calculated PA is also essentially consistent with the
value of $\mathrm{140\pm1.7^{\circ}}$ reported by
\citet{Kraus2012}. We note that the spectroastrometric signature
exhibits no sign of an additional component such as a polar
outflow. This is consistent with the finding of \citet{Kraus2012} who
report that the complex differential phase signatures of this star
are, in-part, the result of the visibility function passing through a
null.

\section{Modelling}
\label{model}

The main aim of this work is to directly constrain the physical
properties and kinematic structure of the disc. We do this by
comparing the spatial and kinematic information contained within the
spectroastrometric data and archival photometric and
spectro-polarimetric data to the state-of-the art radiative transfer
code HDUST developed by \citet{Carciofi2006}. This is done using using
two rotation laws: Keplerian rotation, $v(r) \propto
r^{-\frac{1}{2}}$, and the AMC case, $v(r) \propto r^{-1}$.

\subsection{HDUST:  NLTE radiative transfer}

HDUST is a fully 3D, non-local thermodynamic equilibrium (NLTE), Monte
Carlo code which solves the problems of radiative transfer, radiative
equilibrium and statistical equilibrium for pre-defined gas density
and velocity distributions. Details of the code can be found in
\citet{Carciofi2006,Carciofi2008}. Here, the code is used to generate
the SED, polarisation and H$\alpha$ line profile of a rotating star
surrounded by a gaseous equatorial disc, in addition to making images
of the system. We restrict ourselves to the application of the
code and refer the reader to the references above for more details.

\smallskip

Determining the self consistent structure of Be star discs is a
complex problem, and requires solving the equations for vertical
hydrostatic equilibrium and radial viscous diffusion in non-isothermal
conditions \citep[see][]{Carciofi2008}. This was done by
\citet{Carciofi2008} who demonstrated that in a low density viscous
disc, the temperature structure is approximately isothermal
vertically. Such a disc is well described by a model in which the
vertical density follows a Gaussian distribution and the density
exhibits power-law depencence in the radial direction. $\beta$ CMi
exhibits a low polarisation, indicative of a low density
disc. Consequently, we use this analytical description for the
structure of the disc.

\subsection{Archival data}

Spectro-polarimetric data were obtained using the HPOL
spectropolarimeter, mounted on the Pine Bluff Observatory
telescope. $\beta$ CMi was observed on the night of 22/04/1991 using a
dual Reticon array detector spanning the wavelength range of
3200-7600~$\AA$ and with a spectral resolution of 25~$\AA$. The
Reticon detector was replaced with a CCD detector and two new gratings
which extended the wavelength coverage to 3400-10500$\mathrm{\AA}$ and
improved the spectral resolution to $\mathrm{\sim10\AA}$. $\beta$ CMi
was then observed several times with HPOL (1995 \& 2000). Comparisons
of different observations indicated that the flux blue-wards of the
Balmer jump is possibly affected by systematics. Infrared fluxes were
taken from the 2MASS point source catalogue \citep{Cutri2003}, the
IRAS point source catalogue \citep{IRASPS} and \citep{Ducati2002}. The
UV spectrum was obtained from the INES
database\footnote{http://sdc.laeff.inta.es/ines/} \citep[see][]{ines},
and broad-band measurements of the UV and optical flux were taken from
\citet{Johnson1966,Ducati2002}. Optical spectra were taken from the
HPOL data-base mentioned above.

\section{Results}

\subsection{Modelling procedure and results}

The initial step in reproducing the observations was determining the
properties of the central star. We estimated the radius of the star
and the reddening towards it by reproducing the observed UV and
optical SED with an atmospheric model appropriate for the spectral
type of $\beta$ CMi (B8Ve, $\log = 4.0$,
$T_{\rm{eff}}=12,000~K$). \citet{Fremat2005} report values of $v\sin
i=\mathrm{231~km\,s^{-1}}$ and
$v_{\rm{crit}}=\mathrm{380~km\,s^{-1}}$. Allowing for a ratio
$v_{\rm{rot}}/v_{\rm{crit}}=0.8-0.85$ suggests the inclination is $i
\sim45-50^{\circ}$, in approximate agreement with the results of
\citet{Tycner2005} and \citet{Kraus2012}. We assume an inclination of
$i=40-45^{\circ}$.

\smallskip

It was assumed that the star rotates relatively rapidly, at
approximately 80--85 per cent of its critical velocity. Gravity
darkening effects due to the rapid rotation are taken into account in
HDUST using the von Zeipel flux distribution \citep{vz1924}. The ratio
between the equatorial and polar radii and temperature were determined
based on the stellar rotation and critical velocity. {{To conduct
    the modelling, we explored the available parameter space defined
    by the ranges above. This was done varying $i, V_{\rm{rot}},
    V_{\rm{crit}} \, {\rm{and}} \, R_{\rm{max}}$ by hand as the time
    required to iteratively fit all the observations was
    prohibitively large. $V_{\rm{Turb}} \, {\rm{and}} \,
    \rho_{\rm{0}}$ were treated as free parameters. We note that
    reproducing the photometric, polarimetric and spectroscopic data
    rules out many parameter combinations that might represent local
    minima if fewer types of data were considered.}}

\begin{center}
\begin{table}
\begin{footnotesize}
\begin{center}
\caption{The list of key parameters.\label{pars}}
\begin{tabular}{l l l l}
\hline
Parameter & $\mathrm{Value_{Kep}}$ & $\mathrm{Value_{AMC}}$ &Notes \\
\hline

$d$ (pc)                                                       & 52     & 52     & 1 \\
$i\mathrm{^{\circ}}$                                            & 40     & 45     & 2 \\
$V_{\rm{rot}}/V_{{\rm{crit}}}$                                    & 0.85   & 0.8    & 3 \\
$R_{\rm{e}}$~($\mathrm{R_{\odot}}$)                              & 4.68   & 4.68   & 4 \\
$A_{\rm{V}}$                                                    & 0.05   & 0.05   & 4 \\
$T_{\rm{p}}$(~K)                                                & 15,000 & 15,000 & 4 \\
$R_{\rm{max}}$ (${R_{\star}}$)                                   & 25     & 3.3    & 5 \\
$V_{\rm{Turb}}$ ($\mathrm{km\,s^{-1}}$)                          & 0.1    & 14.5   & 5 \\
$M_{\star}$~($\mathrm{M_{\odot}}$)                               & 3.08   & 3.97   & 6 \\
$R_{\rm{e}}$/$R_{\rm{p}}$                                        & 1.32   & 1.27   & 6 \\
$T_{\rm{p}}$/$T_{\rm{e}}$                                        & 1.48   & 1.37   & 6 \\
$\rho_{\rm{0}}$ ($\mathrm{\times10^{-12}}$$\mathrm{g\,cm^{-3}}$) & 2.9    & 4.6    & 7 \\
$L_{\star}$~($\mathrm{L_{\odot}}$)                               & 220    & 220    & 8 \\
$v \sin i$ ($\mathrm{km\,s^{-1}}$)                              & 182    & 209    & 9 \\
$V_{\rm{crit}} (\mathrm{km\,s^{-1})}$                            & 333    & 370    & 10 \\
\hline 

\end{tabular}
\end{center} {{\bf{Notes:}} 1: $Hipparcos$ parallax
  \citep{Perryman1997}, 2: \citet{Tycner2005} \& \citet{Kraus2012}, 3:
  fit SED and line profile, 4: fit SED, 5: fit line profile, 6: Based
  on $R_p$ and $V_{\rm{crit}}$, 7: fit line flux and polarisation, 8:
  set, 9: c.f. $\mathrm{230\pm16~km\,s^{-1}}$ \citet{Fremat2005} and
  $\mathrm{210 \pm \sim20~km\,s^{-1}}$ \citet{Abt2002}, 10: see e.g. \citet{Fremat2005}.}

\end{footnotesize}
\end{table}
\end{center}

For the Keplerian model, we adopted an inclination of $i = 40^{\circ}$
and rotation at 85 per cent of the critical velocity as this allows a
better match to the observed line profile. This does not produce the
best fit to the HPOL SED, but difference between the predicted SED and
that observed is within the uncertainty in the flux measurements. The
infrared excess of $\beta$ CMi is comparatively small. Reproducing
this requires a low disc density, which is consistent with the low
polarisation of the star. The best fitting density was defined as the
value that reproduced the line-to-continuum ratio of the H$\alpha$
emission. In principle, the central reversal can be recreated by
increasing the line-width. However, this increases the
line-to-continuum ratio. The density cannot be reduced to compensate
without degrading the fit to the IR excess and thus the central
reversal cannot be recreated exactly. We also note that the high
velocity wings of the H$\alpha$ emission cannot be reproduced. This is
attributed to broadening that is not present in the model.

\smallskip

\begin{center}
  \begin{figure}
    \begin{center}
      \begin{tabular}{l}
        {\includegraphics[width=0.33\textwidth]{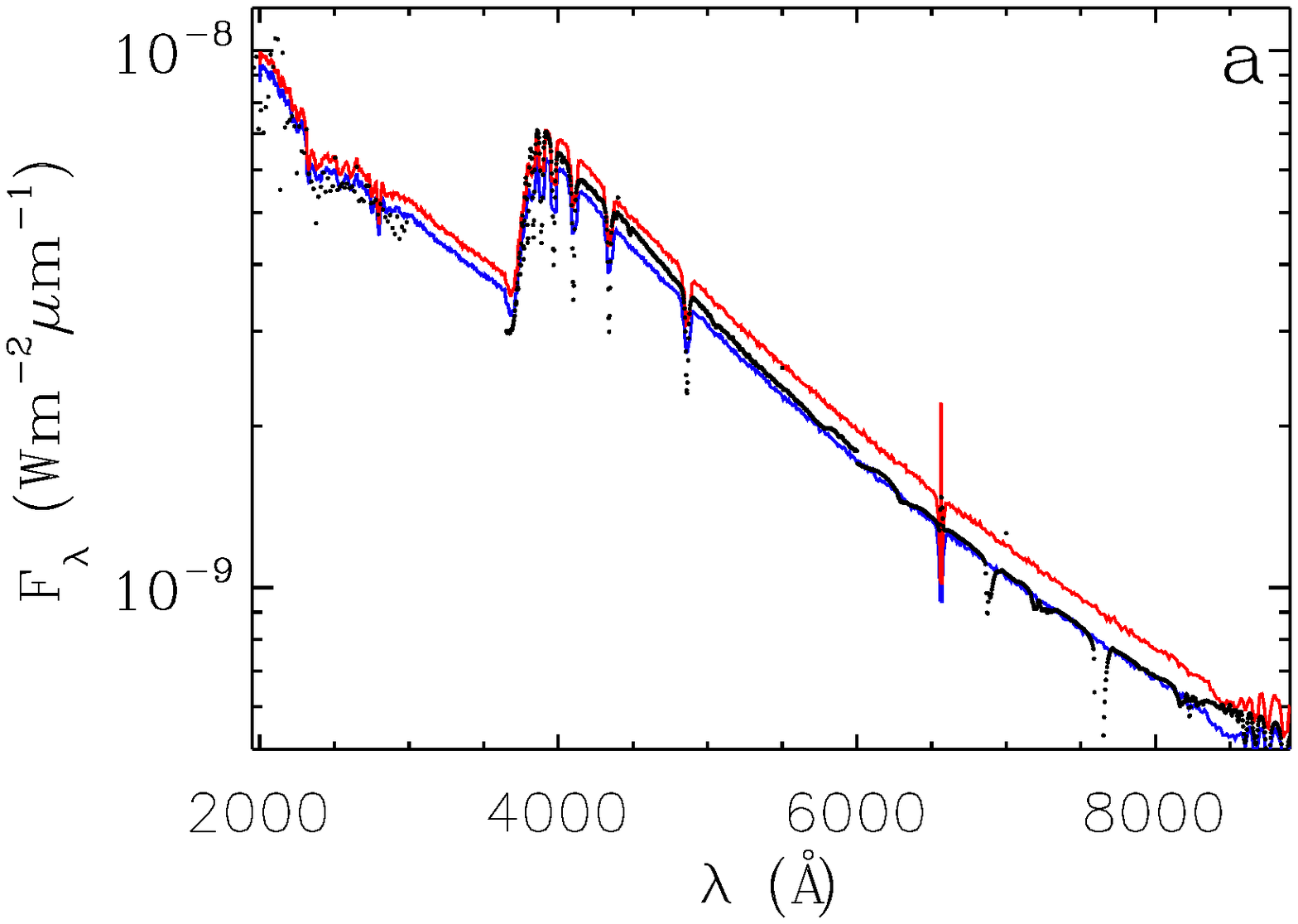}}\\
        {\includegraphics[width=0.33\textwidth]{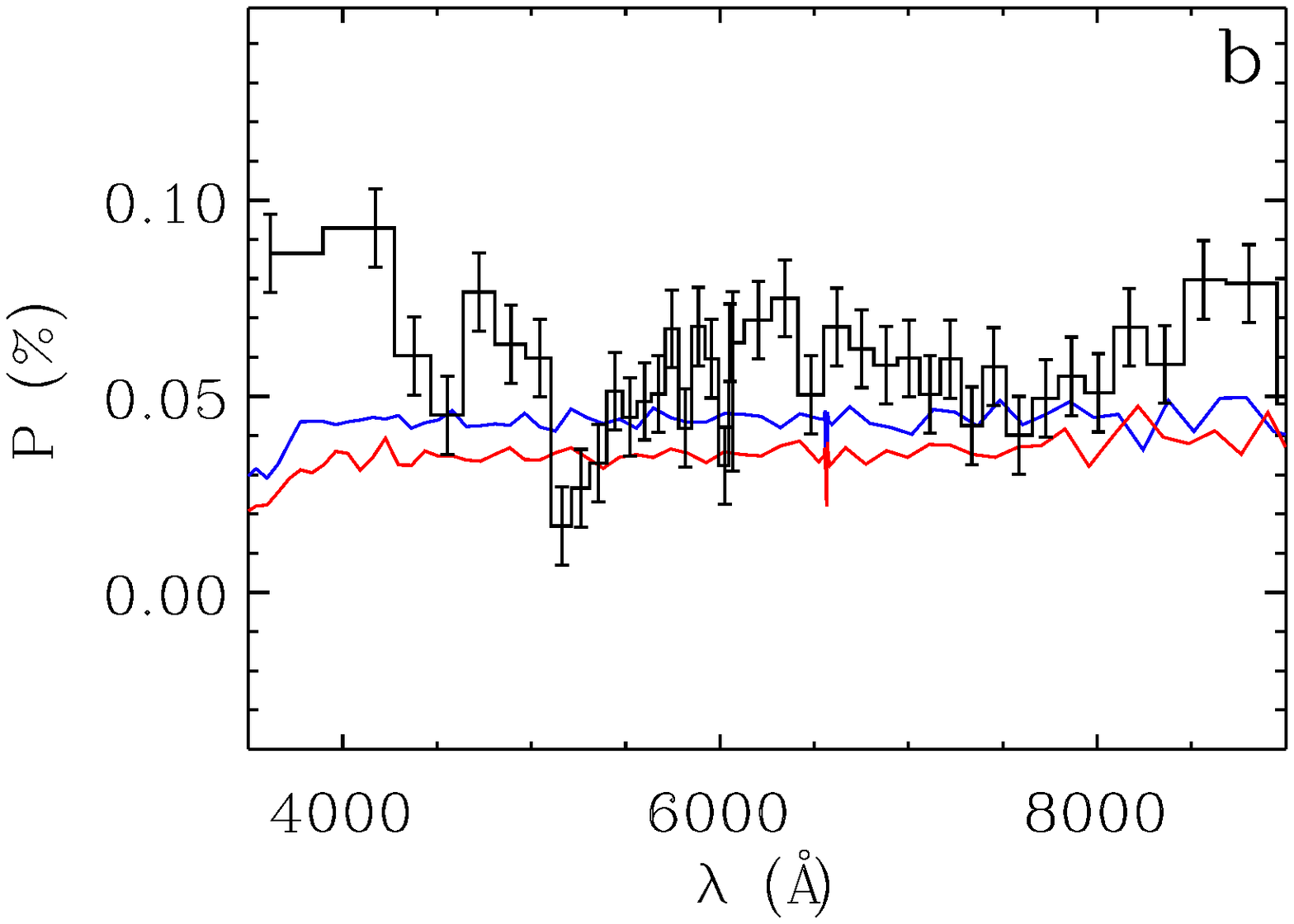}}\\
        {\includegraphics[width=0.33\textwidth]{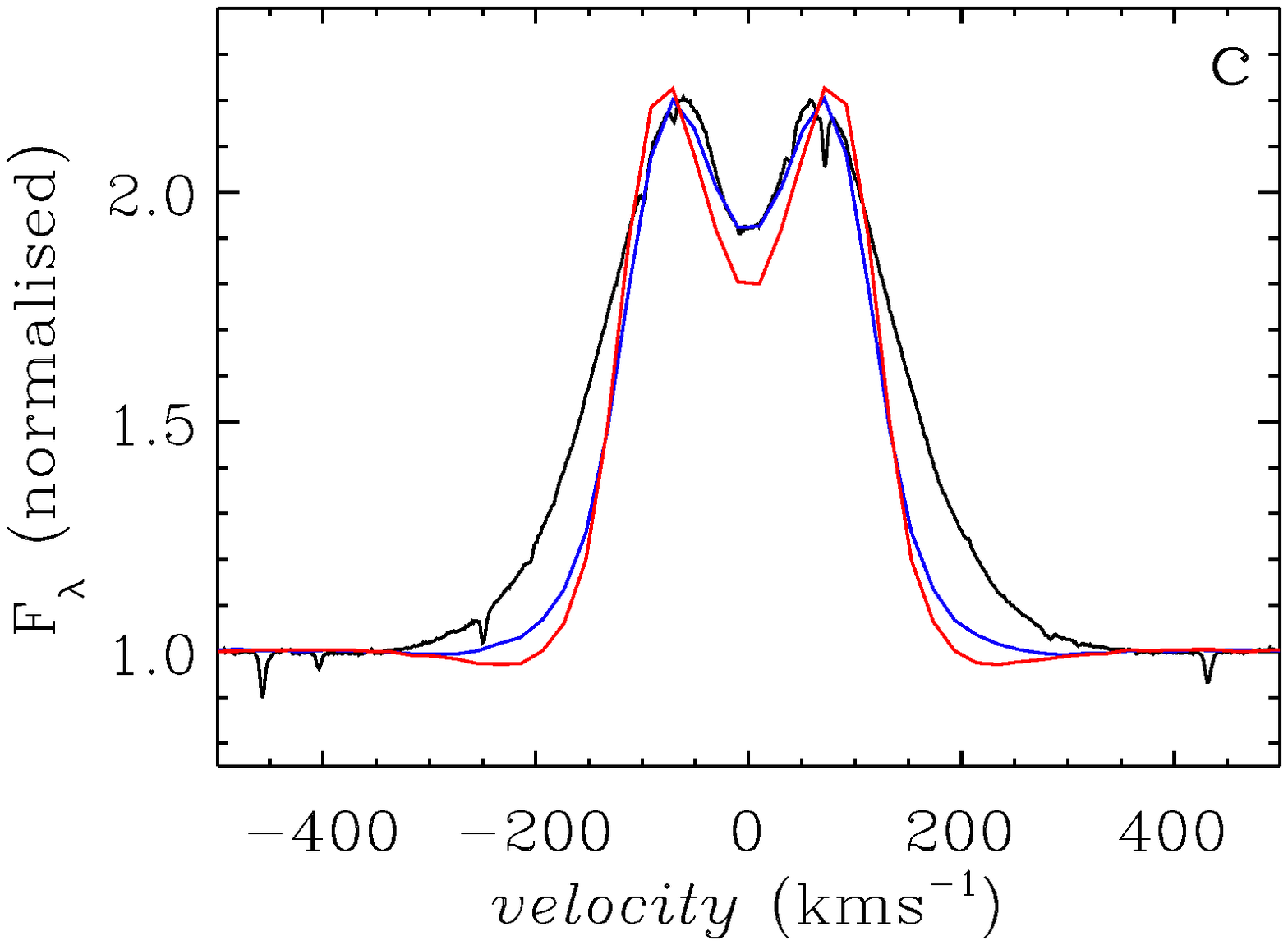}}\\
        {\includegraphics[width=0.35\textwidth]{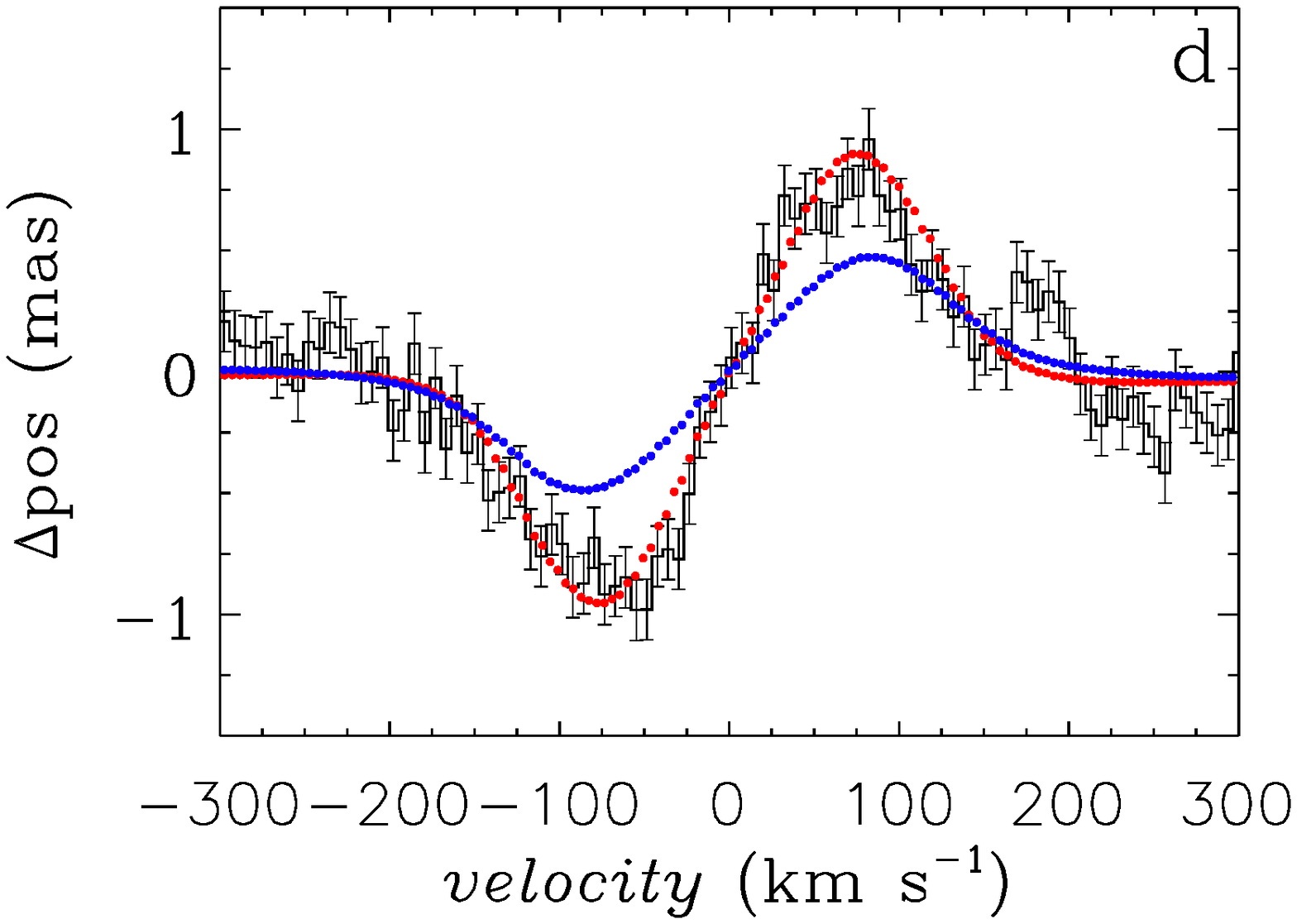}}  \\
      \end{tabular}
      \caption{The final HDUST SEDs (\emph{a}), optical polarisation
        (\emph{b}) and line profiles (\emph{c}) and the predicted
        spectroastrometric signatures (\emph{d}). The observations
        (\emph{black}) are presented alongside the final Keplerian
        (\emph{red}) and AMC (\emph{blue}) models.\label{res_1}}
    \end{center}
  \end{figure}
\end{center}

For the AMC model, we adopted a rotation at 80 per cent of the
critical velocity and an inclination of $i = 45^{\circ}$ as this
allowed the H$\alpha$ line profile to be reproduced. Since the
velocity in the disc decreases more rapidly with radius than in the
Keplerian case, the outer radius of the disc has to be significantly
reduced to recreate the observed double-peaked line profile. To match
the observed line-to-continuum ratio, the line-width also had to be
increased. In this case, this resulted in a good fit to the central
reversal. The final model line profiles, polarisation signatures, SEDs
and spectroastrometric signatures are presented in Figure~\ref{res_1},
and the associated parameters are listed in Table~\ref{pars}. It is
clear that both scenarios are consistent with the entire polarimetric,
photometric (including the IR photometry which is not shown) and
spectroscopic data-set, preventing differentiation between the two
velocity laws on this basis alone.

\subsection{Spectroastrometric results}
The crucial constraint on the disc properties is provided by the
spectroastrometric data. The spectroastrometric signatures of the best
fitting Keplerian and AMC models are presented in panel d) of
Figure~\ref{res_1}. The amplitude of the spectroastrometric signature
of the AMC disc is smaller than that of the Keplerian model; due to
the smaller size of the AMC disc. On the contrary, the signature of
the Keplerian model reproduces the observed signature well ($\chi^2
\sim$1.5 versus $\sim$5). We emphasise that the model astrometric
signatures were predicted by the models that best reproduce the
observed SED, polarisation and line profile. The spectroastrometric
signatures were not fit in any way. {{Since only the Keplerian
    model recreates the observed spectroastrometric signature, the AMC
    scenario can be discarded. Therefore, we confirm that the
    kinematical structure of $\beta$ CMi's disc is consistent with
    Keplerian rotation.}}

\section{Discussion and Conclusion}
\label{discussion}

We present a comparison between photometric, polarimetric and
spectroastrometric observations of the Be star $\beta$ CMi and HDUST,
a NLTE radiative transfer code developed to model Be stars and
discs. Currently, the favoured model of the discs around Be stars is
the viscous decretion disc model. We confront this model with
spectrally resolved sub-mas precision observations which yield direct
evidence of the kinematic structure of the disc around $\beta$ CMi and
provide a unique test of the viscous disc scenario.

\smallskip 

We show that both Keplerian (the prediction of the VDDM) and AMC
scenarios can reproduce the spectroscopic, photometric and
polarimetric data-set. However, we were able to subject the disc
kinematics to a critical test via spectroastrometry which yields
velocity resolved spatial information with sub-mas precision. Due to
the rapid reduction in rotational velocity with radius, the AMC disc
that recreates the observed line profile must be smaller than the
Keplerian disc. As a result, the spectroastrometric signature of the
best fitting AMC disc is smaller than that of the Keplerian
model. {{Consequently, only the Keplerian disc recreates the
spectroastrometric signature observed and the AMC scenario can be
discounted.}} 


\smallskip

That the disc around $\beta$ CMi can be modelled as a viscous disc in
Keplerian rotation indicates that the transport of angular momentum
within it is governed by turbulent viscosity. This addresses one of
the two key questions regarding the discs of Be stars: how is material
injected into the disc and how is angular momentum transported within
it? Although we address the latter question, this does not constrain
the disc formation mechanism. It has been suggested that the disc
around $\beta$ CMi is related to the non-radial pulsations this object
exhibits \citep{Saio2008}. These data cannot directly assess this
possibility. However, our modelling does yield the physical properties of
the disc which may provide useful constraints to future models of disc
formation.

\smallskip

We conclude by emphasising that the comparison between our extensive
data-set and NLTE radiative transfer modelling provides a stringent
test of the properties and kinematics of the disc of $\beta$ CMi. {{The
agreement between the model and observations in both the spectral and
spatial domains clearly demonstrates that the discs around Be stars
can be well modelled as viscous decretion discs.}}

\section*{Acknowledgments}

HEW acknowledges the financial support of the Science and Technology
Facilities Council (STFC) of the UK and the MPIFR. RDO is grateful for
the support of the Leverhulme Trust via the award of a Research
Fellowship.

\bibliographystyle{apj}
\bibliography{/aux/pc20117a/hwwright/Bib/bib}

\label{lastpage}

\end{document}